\documentclass[aps,pre,twocolumn,eqsecnum]{revtex4-2}
\usepackage{graphicx,color}
\usepackage{amsmath}
\usepackage{amssymb}
\usepackage{amsthm}
\usepackage{hyperref}
\usepackage{enumitem}
\usepackage{bm}

\newcommand{\rmi}{\mathrm{i}}
\newcommand{\rme}{\mathrm{e}}

\newcommand{\Tr}{\operatorname{Tr}}

\newcommand{\ket}[1]{|{#1}\rangle}
\newcommand{\bra}[1]{\langle{#1}|}

\newcommand{\ketbras}[3]{\ket{#1}_{#3}\hspace*{-0.2mm}\bra{#2}}

\definecolor{dgreen}{rgb}{0,0.5,0}

\definecolor{dblue}{rgb}{0,0,0.6}

\definecolor{dred}{rgb}{0.784,0,0}

\definecolor{dorange}{cmyk}{0,0.72,1,0.16}
\definecolor{dmagenta}{rgb}{0.847,0.149,0.490}

\definecolor{delete}{cmyk}{0.5,0,0,0}

\begin{document}
\title{Necessity of Feedback Control for the Quantum Maxwell Demon\\in a Finite-Time Steady Feedback Cycle}
\author{Kenta Koshihara}
\affiliation{Department of Physics, Waseda University, Tokyo 169-8555, Japan}
\author{Kazuya Yuasa}
\affiliation{Department of Physics, Waseda University, Tokyo 169-8555, Japan}
\date[]{August 22, 2022}
\begin{abstract}
We revisit quantum Maxwell demon in thermodynamic feedback cycle in the steady-state regime.
We derive a generalized version of the Clausius inequality for a finite-time steady feedback cycle with a single heat bath.
It is shown to be tighter than previously known ones, and allows us to clarify that feedback control is \textit{necessary} to violate the standard Clausius inequality.
\end{abstract}
\maketitle

\section{Introduction}
\label{sec:Introduction}
Over the last decades, our understanding on the role of information in thermodynamics has been greatly advanced~\cite{Maruyama2009,ParrondoHorowitzSagawa2015,Goold2016,Landi2021,Sagawa2022}.
The second law of thermodynamics has been generalized to incorporate the contribution of the information acquired by measurement in thermodynamic processes~\cite{Sagawa2008,Sagawa2009,Jacobs2009,Esposito2017}.
Fluctuation relations for thermodynamic processes involving measurement and feedback have also been explored~\cite{Sagawa2010,Horowitz2010,MorikuniTasaki2011,Sagawa2012a,Sagawa2012b,Funo2013,Campisi2017}.
These developments have been providing us with solid bases to discuss feedback-controlled systems in thermodynamics, including the Maxwell demon~\cite{Maruyama2009}.

The renewed interests in thermodynamics extend to the quantum domain~\cite{QuantumThermodynamics}.
In quantum thermodynamics, various quantum features, such as quantum coherence, quantum entanglement, and uncertainty principle, can play roles and can give rise to thermodynamic effects that are absent in the classical regime~\cite{Brandner2015,Francica2017,Francica2020}.
Among various features, we wish to look into the effects of quantum measurement in thermodynamics.
Quantum measurement disturbs the state of the measured system.
This effect should be properly taken into account in quantum thermodynamics.
Moreover, by this effect, quantum measurement can extract/inject energy from/to a quantum working substance~\cite{Campisi2019}.
In other words, quantum measurement in a quantum thermodynamic cycle should be counted as a thermodynamic ``stroke.''
An extreme idea in this direction leads to quantum engines driven by quantum measurements~\cite{Elouard2017a,Elouard2017b,YiTalkerKim2017,Anders2017,Yi2017,Elouard2018,YiTalkerKim2018,Buffoni2019,Seah2020,Behzadi2021,Bresque2021,Anka2021,Lin2021,Manikandan2022,Lisboa2022}, where quantum measurement acts as a heat bath of a heat engine, fueling energy to its working substance.

In this paper, we would like to discuss the following question: \textit{Is feedback control really necessary to realize quantum Maxwell demon?}
If the result obtained in the previous works~\cite{Sagawa2008,Jacobs2009,Funo2013} is applied to a feedback cycle with a single heat bath, a version of generalized second law of thermodynamics
\begin{equation}
\beta Q\le I_\mathrm{QC}
\label{eq:SecondLaw0}
\end{equation}
is obtained for the heat $Q$ extracted from the heat bath at an inverse temperature $\beta$ per cycle.
The quantity $I_\mathrm{QC}$ on the right-hand side is called QC-mutual information, and quantifies the amount of information acquired by the measurement performed in the feedback cycle.
Its definition is found also in Sec.~\ref{sec:FirstSecondLaws} of this paper.
This inequality~(\ref{eq:SecondLaw0}) generalizes the standard Clausius inequality $\beta Q\le0$, and reveals that, if one acquires information $I_\mathrm{QC}>0$ by the measurement, there is a chance to get a positive heat $Q>0$ after a cycle, violating the standard Clausius inequality $\beta Q\le0$.
It is said that one can violate the standard Clausius inequality $\beta Q\le0$ by exploiting the information $I_\mathrm{QC}>0$ by feedback control depending on the outcome of the measurement.
In the quantum case, however, it is not obvious, since the backaction of the quantum measurement can supply/dissipate the energy of the working substance, and might result in an incoming heat $Q>0$ from the heat bath after a cycle, even without performing any feedback.
The generalized Clausius inequality~(\ref{eq:SecondLaw0}) is not informative enough to clarify this point.

There is actually an answer to this question in the quantum case.
It is shown in Ref.~\cite{MorikuniTasaki2011} that the fluctuation relation derived there reduces to the quantum Jarzynski equality in the absence of feedback, if the quantum measurement performed in the cycle is unital (we will be interested in ``bare''~\cite{Jacobs2009,JacobsBook2014} or ``minimally disturbing''~\cite{WisemanMilburnBook2010} quantum measurements, which are unital measurements, for the reason recalled in Sec.~\ref{sec:Measurement}).
Then, using this fact (and closing the cycle by adding an additional thermalization step to the protocol considered in Ref.~\cite{MorikuniTasaki2011}), one can show that the standard Clausius inequality $\beta Q\le0$ holds for the cycle without feedback control.
Feedback control is thus \textit{necessary} to violate the standard Clausius inequality $\beta Q\le0$.

\begin{figure*}
\includegraphics[scale=0.5]{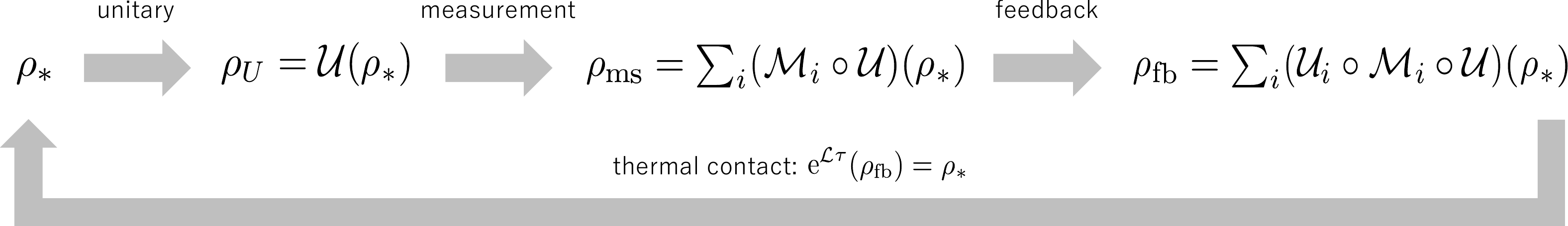}
\caption{The thermodynamic cycle studied in this paper. We will analyze thermodynamic quantities in the steady cycle, where $\rho_*$ is the fixed point of the map $\mathcal{E}=\sum_i\rme^{\mathcal{L}\tau}\circ\mathcal{U}_i\circ\mathcal{M}_i\circ\mathcal{U}$ for a finite time $\tau$, satisfying $\mathcal{E}(\rho_*)=\rho_*$. The fixed point $\rho_*$ generally differs from the thermal equilibrium state $\rho_\mathrm{th}$ for a finite $\tau$, even though the evolution $\rme^{\mathcal{L}\tau}$ describing the heat exchange with the heat bath is assumed to thermalize the system $\rme^{\mathcal{L}\tau}(\rho)\to\rho_\mathrm{th}$ in the long-time limit $\tau\to\infty$.}
\label{fig:Cycle}
\end{figure*}

The fluctuation relation derived in Ref.~\cite{MorikuniTasaki2011}, however, requires that the protocol starts with the thermal equilibrium state $\rho_\mathrm{th}$ at the inverse temperature $\beta$.
If a cycle starts and ends with the thermal equilibrium state $\rho_\mathrm{th}$, the above conclusion can be proven in a more direct way, on the basis of the passivity of the thermal equilibrium state $\rho_\mathrm{th}$ (see the discussion in Sec.~\ref{sec:Example} of this paper; see also Ref.~\cite{Anka2021} as well as Refs.~\cite{YiTalkerKim2017,Lin2021}).
The question we wish to ask in this paper is actually the following one: \textit{Is feedback control necessary to realize quantum Maxwell demon in finite-time steady cycles}?
We consider a feedback cycle which closes with the thermal contact with a heat bath only for a finite time.
As this cycle is repeated, the evolution of the working substance approaches a steady cycle.
Since the thermal contact for a finite time does not completely thermalize the working substance, the steady cycle starts and ends with a nonthermalized state $\rho_*$.
The passivity of the initial state is lost, the above passivity argument does not apply, and there might be a possibility that the standard Clausius inequality $\beta Q\le0$ would be violated solely by the backaction of the measurement without any feedback in the finite-time steady cycles.

The generalized second law~(\ref{eq:SecondLaw0}), which is too originally derived for a protocol starting with the thermal equilibrium state $\rho_\mathrm{th}$~\cite{Sagawa2008,Jacobs2009,Funo2013}, can be generalized to such finite-time steady cycles, but it remains uninformative to answer the question.
On the other hand, in this paper, we are going to derive an improved Clausius inequality valid for a finite-time steady cycle, which is tighter than the inequality~(\ref{eq:SecondLaw0}), and give the answer to the question: \textit{Feedback control is necessary to realize quantum Maxwell demon even in finite-time steady feedback cycles}.

The paper is organized as follows.
We start by presenting the thermodynamic feedback cycle considered in this work in Sec.~\ref{sec:Setup}\@.
Since the structure of quantum measurement is important for our discussion, we recall it in Sec.~\ref{sec:Measurement}\@.
We then introduce the relevant thermodynamic quantities and derive the first and second laws of thermodynamics for the steady feedback cycle in Sec.~\ref{sec:FirstSecondLaws}\@.
In particular, we present an improved generalized Clausius inequality, which holds for finite-time steady cycles, is tighter than previously known bounds, and allows us to draw the answer to the above question in Sec.~\ref{sec:Necessity}\@.
The performance of the steady feedback cycle is numerically demonstrated with a two-level working substance in Sec.~\ref{sec:Example}, and we conclude the paper in Sec.~\ref{sec:Conclusions}\@.
The derivation of the improved Clausius inequality is presented in Appendix~\ref{appendix:proof2ndlaw}, and a proof of the passivity of the thermal equilibrium state against unital measurements is provided in Appendix~\ref{appendix:backaction}\@.

\section{Thermodynamic Feedback Cycle}
\label{sec:Setup}
We consider the following thermodynamic feedback cycle. 
Consider a quantum system $S$, whose initial Hamiltonian is given by $H_S$, and a heat bath $B$ at a temperature $T$.
Then, the protocol proceeds as follows.
\begin{enumerate}
\item
We first apply a unitary control $\mathcal{U}$ on $S$, and the state of $S$ is transformed from $\rho_\mathrm{ini}$ to $\rho_U=\mathcal{U}(\rho_\mathrm{ini})$.

\item
We perform measurement on $S$ and get a measurement outcome $i$ with probability $p_i$. 
The state of $S$ is changed from $\rho_U$ to $\rho_\mathrm{ms}^{(i)}=\mathcal{M}_i(\rho_U)/p_i$.

\item
We apply a unitary feedback $\mathcal{U}_i$ on $S$ depending on the outcome $i$ of the measurement.
The state of $S$ becomes $\rho_\mathrm{fb}^{(i)}=\mathcal{U}_i(\rho_\mathrm{ms}^{(i)})$.

\item
Finally, we put $S$ in contact with the heat bath $B$ for time $\tau$, to bring $S$ towards the thermal equilibrium state at the temperature $T$. We describe this process by a Markovian generator $\mathcal{L}$, and $S$ evolves to $\rho_\mathrm{fin}^{(i)}=\rme^{\mathcal{L}\tau}(\rho_\mathrm{fb}^{(i)})$.
\end{enumerate}
We repeat this cycle many times, and analyze the behaviors of thermodynamic quantities in steady cycles.
See Fig.~\ref{fig:Cycle}.

The unitary control in Step~1 is represented by $\mathcal{U}(\rho_\mathrm{ini})=U\rho_\mathrm{ini}U^\dag$ with a unitary operator $U$.
It is realized by driving the Hamiltonian of $S$.

The measurement process in Step~2 is described by a set of completely positive (CP) linear maps $\{\mathcal{M}_i\}$, which is called CP instrument~\cite{Hayashi2015}.
The probability of obtaining measurement outcome $i$ is given by the normalization factor $p_i=\Tr\mathcal{M}_i(\rho_U)$, and the sum over all possible outcomes $\mathcal{M}=\sum_i\mathcal{M}_i$ is a completely positive and trace-preserving (CPTP) map.
The structure of the maps $\{\mathcal{M}_i\}$ is important in the following discussion, and is recapitulated in Sec.~\ref{sec:Measurement}\@.

The feedback controls in Step~3 are represented by $\mathcal{U}_i(\rho_\mathrm{ms}^{(i)})=U_i\rho_\mathrm{ms}^{(i)}U_i^\dag$ with unitary operators $U_i$.
They are realized by driving the Hamiltonian of $S$ depending on the outcome $i$ of the measurement.
We assume that the Hamiltonian is back to the initial one $H_S$ at the end of this driving and is kept during Step~4, so that the cycle is completed after Step~4.

The thermalization process in Step~4 is assumed to be described by a Markovian master equation, with a generator $\mathcal{L}$ of the Gorini-Kossakowski-Lindblad-Sudarshan (GKLS) form~\cite{ref:DynamicalMap-Alicki,ref:GKLS-DariuszSaverio}.
Its explicit form is not important, but it has to admit the thermal equilibrium state $\rho_\mathrm{th}=\rme^{-\beta H_S}/Z_S$ with $Z_S=\Tr\rme^{-\beta H_S}$ as its unique stationary state, satisfying $\mathcal{L}(\rho_\mathrm{th})=0$, where $\beta=(k_BT)^{-1}$ is the inverse temperature of the heat bath $B$ with $k_B$ the Boltzmann constant.
One can think of a standard amplitude-damping channel obeying the detailed balance condition, such as the one considered in Sec.~\ref{sec:Example}\@.

The outcome $i$ of the measurement appears just probabilistically, with probability $p_i$, and the state $\rho_\mathrm{fin}^{(i)}$ after a cycle depends on the outcome of the measurement.
We consider the average state $\rho_\mathrm{fin}=\sum_ip_i\rho_\mathrm{fin}^{(i)}$ over all possible outcomes of the measurement.
The average evolution of $S$ by the cycle is given by the CPTP map $\mathcal{E}=\sum_i\rme^{\mathcal{L}\tau}\circ\mathcal{U}_i\circ\mathcal{M}_i\circ\mathcal{U}$.
If this map $\mathcal{E}$ is mixing, i.e., if $\mathcal{E}^N(\rho)\to\rho_*$ as $N\to\infty$, converging to a unique fixed point $\rho_*$ for any input state $\rho$~\cite{Burgarth2007,ref:ConvexErgodicity}, the evolution of $S$ approaches a limit cycle.
We are going to discuss thermodynamics in such steady cycles with finite $\tau$~\cite{Esposito2017}, in which $\rho_\mathrm{ini}=\rho_\mathrm{fin}=\rho_*$.
Note that for $\tau\to\infty$ system $S$ completely thermalizes to $\rho_\mathrm{th}$ after every cycle, and we have $\rho_*=\rho_\mathrm{th}$.
If $\tau$ is finite, on the other hand, system $S$ does not completely thermalize to $\rho_\mathrm{th}$ after a cycle, but returns to a nontrivial $\rho_*$ in the steady cycle.
With a finite $\tau$, we are allowed to discuss ``powers'' of the thermodynamic cycle such as the heat flow and the work extraction per unit time~\cite{Shiraishi2016,Shiraishi2017,Abiuso2019,Vittorio2019,Uchiyama2021,Vittorio2021}.

There are minor differences among the protocols considered in the previous works.
For instance, the protocols studied in Refs.~\cite{Sagawa2008,Jacobs2009,YiTalkerKim2017,Yi2017,YiTalkerKim2018,Behzadi2021,Anka2021,Lin2021,Lisboa2022} consist basically of the same steps as the one considered here, but they are assumed to start and end with the thermal equilibrium state.
In addition, in Ref.~\cite{Sagawa2008}, the driven Hamiltonian need not return to the initial one at the end of the protocol.
In the protocol considered here, on the other hand, the driven Hamiltonian gets back to the initial one $H_S$ after Step~3 so that the cycle is closed after Step~4, and the cycle is repeated with a finite time $\tau$ without waiting for the thermalization of the working substance $S$ in Step~4.
Reference~\cite{Esposito2017} studies a finite-time steady cycle, but the protocol does not have Step~1 (i.e.~$U=\openone$, or it is absorbed in Step~2), and the exchange of heat with a heat bath occurs during the feedback control.
Heat exchange is allowed also during other steps in the protocol of Ref.~\cite{Sagawa2008}.
In our protocol, the feedback process in Step~3 and the thermal contact in Step~4 are separated.
The protocol considered in Ref.~\cite{MorikuniTasaki2011} to develop fluctuation relations starts with the thermal equilibrium state and ends without the last thermalization process, but some of the results obtained there can be compared with the present work by adding the thermalization step to close the cycle.
Finally, we assume that the thermal contact is described by a Markovian generator, while it is described by a unitary process acting on the coupled system $S+B$ in Refs.~\cite{Sagawa2008,Esposito2017}.
In other words, we assume weak interaction between system $S$ and heat bath $B$ in Step~4 of the protocol considered here.
It is also the case in Refs.~\cite{Jacobs2009,YiTalkerKim2017,Yi2017,YiTalkerKim2018,Behzadi2021,Anka2021,Lin2021,Lisboa2022}, where system $S$ is assumed to be thermalized with negligible correlations with heat bath $B$ after thermalization.

\section{Pure Quantum Measurement}
\label{sec:Measurement}
Before starting to discuss thermodynamics in the cycle introduced in Sec.~\ref{sec:Setup}, let us recall how to describe general quantum measurements.
We stress that \textit{general quantum measurements implicitly include feedback controls in their structure}~\cite{JacobsBook2014}. 
It is important to identify which part of the effect of a general quantum measurement is considered to be purely due to measurement and which part should be counted as feedback control.
Such classification between measurements and feedback controls allows us to clarify the essential roles of quantum measurements and feedback controls in the thermodynamic cycle.

The statistics of the outcomes $\{i\}$ of a general quantum measurement is characterized by a positive operator-valued measure (POVM) $\{\Pi_i\}$ with $\Pi_i\ge0$ and $\sum_i\Pi_i=\openone$~\cite{Nielsen2000,WisemanMilburnBook2010,JacobsBook2014,Hayashi2015}.
The probability of getting outcome $i$ by the measurement in a state $\rho$ is given by $p_i=\Tr(\Pi_i\rho)$. 
On the other hand, the disturbance on the system by the measurement is described by a set of CP maps $\{\mathcal{M}_i\}$, called CP instrument~\cite{Hayashi2015}.
When outcome $i$ is obtained by the measurement, the state of the measured system is changed as $\rho\to\mathcal{M}_i(\rho)$ apart from the normalization. 
The normalization gives the probability of obtaining the outcome $i$, i.e., $p_i=\Tr\mathcal{M}_i(\rho)=\Tr(\Pi_i\rho)$, and the sum over all possible outcomes of the measurement $\mathcal{M}=\sum_i\mathcal{M}_i$ is CPTP\@.
The CP instrument of a general quantum measurement reads~\cite{WisemanMilburnBook2010,JacobsBook2014,Hayashi2015}
\begin{equation}
\mathcal{M}_i(\rho)
=
\sum_\alpha K_i^{(\alpha)}\rho K_i^{(\alpha)\dagger}.
\label{eq:GeneralMeas}
\end{equation}
In this case, the POVM element for the outcome $i$ of the measurement is given by $\Pi_i=\sum_\alpha K_i^{(\alpha)\dagger}K_i^{(\alpha)}$ in terms of the measurement operators $\{K_i^{(\alpha)}\}$ satisfying $\sum_i\sum_\alpha K_i^{(\alpha)\dagger}K_i^{(\alpha)}=\openone$.

Note that the measurement described by a CP map $\mathcal{M}_i$ of the form~(\ref{eq:GeneralMeas}) can be interpreted as a ``coarse-grained measurement,'' or an ``inefficient measurement''~\cite{Jacobs2006,Jacobs2009,WisemanMilburnBook2010,Funo2013,JacobsBook2014}.
Indeed, consider a measurement yielding pairs of measurement outcomes $(i,\alpha)$, but discard the outcomes $\alpha$, keeping only outcomes $i$.
This is a nonselective measurement regarding $\alpha$, and this feature is represented by the summation over $\alpha$ in~(\ref{eq:GeneralMeas}).
In contrast, if each element $\mathcal{M}_i$ of a CP instrument consists only of a single measurement operator $K_i$ as
\begin{equation}
\mathcal{M}_i^\mathrm{E}(\rho)
=
K_i\rho K_i^\dagger,
\label{eq:EfficientMeas}
\end{equation}
it would be considered as an ``efficient measurement,'' in the sense that no coarse-graining is involved.
In this case, the corresponding POVM element reads $\Pi_i=K_i^\dagger K_i$, and $\sum_iK_i^\dag K_i=\openone$.

Those ``measurements,'' however, would not be simply considered as measurements, but could be regarded as quantum operations involving feedback controls.
To see this, consider the polar decomposition of each measurement operator $K_i$ of the efficient measurement~(\ref{eq:EfficientMeas}),
\begin{equation}
K_i=V_iM_i,
\quad
M_i\geq0,
\label{eq:polardecomp}
\end{equation}
where $M_i$ is a positive-semidefinite operator and $V_i$ is a unitary~\cite{Nielsen2000,WisemanMilburnBook2010,HornJohnson2012,JacobsBook2014,Hayashi2015}.
\textit{This unitary $V_i$ can be considered as an operation applied depending on the outcome $i$ of the measurement represented by the measurement operators $\{M_i\}$}~\cite{JacobsBook2014}.
The measurement process~(\ref{eq:EfficientMeas}) is indistinguishable with the sequence of operations where first the measurement with $\{M_i\}$ is performed and then the feedback control is applied with $\{V_i\}$.
The measurement
\begin{equation}
\mathcal{M}_i^\mathrm{B}(\rho)
=
M_i\rho M_i,
\quad
M_i\geq0,
\label{eq:PureMeas}
\end{equation}
with the unitaries $\{V_i\}$ removed from the measurement operators, is called ``bare measurement''~\cite{Jacobs2009,JacobsBook2014} or ``minimally disturbing measurement''~\cite{WisemanMilburnBook2010}, and can be regarded as ``pure quantum measurement.''
Note that the set of operators $\{M_i\}$ extracted by the polar decomposition in~(\ref{eq:polardecomp}) satisfies the condition for a CP instrument, $\sum_iM_i^2=\sum_iK_i^\dag K_i=\openone$.
In this case, the POVM elements are given by $\Pi_i=M_i^2$.
Note that this bare quantum measurement $\{\mathcal{M}_i^\mathrm{B}\}$ is a kind of ``unital quantum measurement'' $\{\mathcal{M}_i^\mathrm{U}\}$ whose map $\mathcal{M}^\mathrm{U}=\sum_i\mathcal{M}_i^\mathrm{U}$ is unital, satisfying $\mathcal{M}^\mathrm{U}(\openone)=\openone$.

The general quantum measurement $\{\mathcal{M}_i\}$ in~(\ref{eq:GeneralMeas}) is also regarded as quantum operation consisting of bare quantum measurement and feedback control.
The polar decomposition $K_i^{(\alpha)}=V_i^{(\alpha)}M_i^{(\alpha)}$ with $M_i^{(\alpha)}$ positive-semidefinite and $V_i^{(\alpha)}$ unitary allows us to interpret the process represented by the CP map~(\ref{eq:GeneralMeas}) as the process where first efficient bare measurement with $\{M_i^{(\alpha)}\}$ is performed and then feedback control is applied with $\{V_i^{(\alpha)}\}$ depending on the outcome $(i,\alpha)$ of the bare measurement.
In this way, the general quantum measurement $\{\mathcal{M}_i\}$ in~(\ref{eq:GeneralMeas}) is endowed with the feedback structure.

The separation of feedback control and bare measurement helps us identify the effects which are purely due to measurements and those which can be considered due to feedback controls in the thermodynamic cycle.

\section{Thermodynamics}
Let us now discuss thermodynamics in the feedback cycle introduced in Sec.~\ref{sec:Setup}\@.

\subsection{First and Second Laws}
\label{sec:FirstSecondLaws}
The unitary operation in Step~1 of the cycle changes the energy of system $S$ by $W_U=\Tr(H_S\rho_U)-\Tr(H_S\rho_\mathrm{ini})$, which is considered as work done on $S$~\cite{note:2}.
The quantum measurement performed in Step~2 disturbs the state of $S$, and the energy of $S$ is changed again.
The amount of the change in the energy of $S$ by the quantum measurement is given by $E_\mathrm{ms}=\sum_ip_i\Tr(H_S\rho_\mathrm{ms}^{(i)})-\Tr(H_S\rho_U)$. 
The change in the energy of $S$ by the feedback control in Step~3 is considered as additional work done on $S$, which reads $W_\mathrm{fb}=\sum_ip_i\Tr(H_S\rho_\mathrm{fb}^{(i)})-\sum_ip_i\Tr(H_S\rho_\mathrm{ms}^{(i)})$.
The energy transferred from heat bath $B$ to system $S$ in Step~4 is heat.
Under the assumption of the Markovianity of the thermalization process (with weak interactions with the heat bath), it is estimated by $Q=\sum_ip_i\Tr(H_S\rho_\mathrm{fin}^{(i)})-\sum_ip_i\Tr(H_S\rho_\mathrm{fb}^{(i)})$.  
In the steady cycle, where $\rho_\mathrm{ini}=\rho_\mathrm{fin}=\rho_*$, we get
\begin{equation}
W_U+E_\mathrm{ms}+W_\mathrm{fb}+Q=0.
\label{eq:FirstLaw}
\end{equation}
This is the first law of thermodynamics in the protocol considered here.

\begin{figure*}
\centering
\includegraphics[scale=0.45]{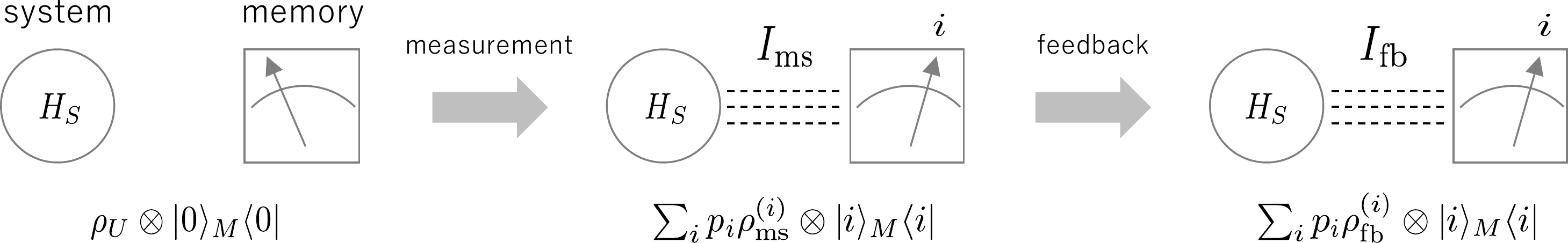}
\caption{The quantities $I_\mathrm{ms}$ and $I_\mathrm{fb}$ appearing in the inequality~(\ref{eq:SecondLaw1}) can be interpreted as the correlations after the quantum measurement and after the feedback control, respectively, between system $S$ and memory $M$ recording the outcome of the measurement. 
Suppose that memory $M$ is initially prepared in a state, say $\ketbras{0}{0}{M}$, and after the quantum measurement the outcome $i$ of the measurement is recorded in a state $\ketbras{i}{i}{M}$ of a complete set of orthonormal basis states of memory $M$.
A feedback control $U_i$ is then applied depending on the outcome $i$ of the measurement.
In this scenario, the state, averaged over all possible outcomes $i$, of the composite system $S+M$ after the measurement (feedback) is given by 
$\rho_{SM}^{(\mathrm{ms(fb)})}=\sum_ip_i\rho_\mathrm{ms(fb)}^{(i)}\otimes\ketbras{i}{i}{M}$.
The quantum mutual information between $S$ and $M$ in this state gives $I(\rho_{SM}^{(\mathrm{ms(fb)})})=I_\mathrm{ms(fb)}$ appearing in the inequality~(\ref{eq:SecondLaw1}), where the quantum mutual information is defined by $I(\rho_{SM})=D(\rho_{SM}\|\rho_S\otimes\rho_M)=S(\rho_S)+S(\rho_M)-S(\rho_{SM})$ with $\rho_S=\Tr_M\rho_{SM}$ and $\rho_M=\Tr_S\rho_{SM}$, and quantifies the correlation between $S$ and $M$~\cite{Nielsen2000,Hayashi2015,HolevoBook2019,Landi2021,Sagawa2022}. Here, $D(\rho\|\sigma)=\Tr[\rho(\log\rho-\log\sigma)]$ is the quantum relative entropy and $S(\rho)=-\Tr(\rho\log\rho)$ is the von Neumann entropy. Note that the quantum mutual information is non-negative $I(\rho_{SM})\ge0$, since the quantum relative entropy is non-negative $D(\rho\|\sigma)\ge0$ for any density operators $\rho$ and $\sigma$.}
\label{fig:DemonMemory}
\end{figure*}

As a second law of thermodynamics, we here present the following generalized version of Clausius inequality for the steady cycle:
\begin{equation}
\beta Q
\leq
-(\Delta S_\mathrm{ms}+\Delta I_\mathrm{fb}).
\label{eq:SecondLaw1}
\end{equation}
This is the key formula of the present paper.
Note that this holds for the finite-time steady cycle.
This simply follows from the monotonicity of the relative entropy under the action of any CPTP map~\cite{Hayashi2015,HolevoBook2019,Landi2021,Sagawa2022} (under the action of $\rme^{\mathcal{L}\tau}$ in the present case).
See Appendix~\ref{appendix:proof2ndlaw} for the derivation of~(\ref{eq:SecondLaw1}).
In this inequality, $\Delta S_\mathrm{ms}=S(\sum_ip_i\rho_\mathrm{ms}^{(i)})-S(\rho_U)$ represents the increase in the von Neumann entropy $S(\rho)=-\Tr(\rho\log\rho)$ of system $S$ by the quantum measurement performed in Step~2 of the steady cycle.
The other quantity $\Delta I_\mathrm{fb}=I_\mathrm{fb}-I_\mathrm{ms}$ is defined in terms of $I_\mathrm{ms(fb)}=S(\sum_ip_i\rho_\mathrm{ms(fb)}^{(i)})-\sum_ip_iS(\rho_\mathrm{ms(fb)}^{(i)})$, which can be interpreted as the correlation after the measurement (feedback) between system $S$ and memory $M$ storing the outcome $i$ of the measurement, and is non-negative $I_\mathrm{ms(fb)}\ge0$.
See Fig.~\ref{fig:DemonMemory}.
The quantity $\Delta I_\mathrm{fb}$ represents the change in the correlation by the feedback control.

If the right-hand side of the inequality~(\ref{eq:SecondLaw1}) is vanishing (or negative), we have $\beta Q\le0$, which is the standard Clausius inequality.
However, due to the backaction of the quantum measurement $\Delta S_\mathrm{ms}$ and the effect of the feedback control $\Delta I_\mathrm{fb}$, the right-hand side of the inequality~(\ref{eq:SecondLaw1}) can be positive, and the heat $Q$ can actually become positive, leading to the violation of the standard Clausius inequality (see the example studied in Sec.~\ref{sec:Example}).

If we apply the inequality obtained in the previous works~\cite{Sagawa2008,Jacobs2009,Funo2013} to the present scenario, we get the inequality~(\ref{eq:SecondLaw0}), i.e.
\begin{equation}
\beta Q
\le
I_\mathrm{QC},
\label{eq:SecondLaw2}
\end{equation}
where $I_\mathrm{QC}=S(\rho_U)-\sum_ip_iS(\rho_\mathrm{ms}^{(i)})$ is the QC-mutual information, which quantifies the amount of information acquired by the quantum measurement. 
The inequality~(\ref{eq:SecondLaw1}) derived here is tighter than this inequality~(\ref{eq:SecondLaw2}).
Indeed, the inequality~(\ref{eq:SecondLaw1}) is bounded by
\begin{align}
\beta Q&\le-(\Delta S_\mathrm{ms}+\Delta I_\mathrm{fb})
\nonumber\\
&
=I_\mathrm{QC}-I_\mathrm{fb}
\nonumber\\
&\le I_\mathrm{QC},
\label{eq:SecondLaw12}
\end{align}
since $I_\mathrm{fb}\ge0$ as mentioned above.
This also shows that the inequality~(\ref{eq:SecondLaw2}) holds also for the finite-time steady cycle, while it was originally derived in Refs.~\cite{Sagawa2008,Jacobs2009,Funo2013} for a protocol which starts with the thermal equilibrium state $\rho_\mathrm{th}$.

If we apply the formalism developed in Ref.~\cite{Esposito2017} to the present scenario, a different inequality is obtained,
\begin{equation}
\beta Q\le\Delta S_M,
\label{eq:SecondLaw3}
\end{equation}
where $\Delta S_M=S(\rho_M^\mathrm{(ms)})-S(\rho_M)$ represents the change in the von Neumann entropy of the state of memory $M$ by the measurement process.
Note that the memory scheme considered in Ref.~\cite{Esposito2017} is different from the one considered in Fig.~\ref{fig:DemonMemory}.
In Ref.~\cite{Esposito2017}, the measurement process is described by a unitary transformation $U_\mathrm{ms}$ acting on the composite system $S+M$, and the feedback control is represented by a unitary gate $U_\mathrm{fb}$ on $S+M$ controlled by the state of memory $M$.
The state of $S+M$ after the measurement is given by $\rho_{SM}^{\mathrm{(ms)}}=U_\mathrm{ms}(\rho_U\otimes\rho_M)U_\mathrm{ms}^\dag$, and it is further transformed to $\rho_{SM}^{\mathrm{(fb)}}=U_\mathrm{fb}U_\mathrm{ms}(\rho_U\otimes\rho_M)U_\mathrm{ms}^\dag U_\mathrm{fb}^\dag$ by the feedback control.
The state of memory $M$ after these processes is given by $\rho_M^\mathrm{(ms)}=\Tr_S\rho_{SM}^\mathrm{(ms)}=\Tr_S\rho_{SM}^\mathrm{(fb)}$.
The inequality~(\ref{eq:SecondLaw1}) derived here is also tighter than the inequality~(\ref{eq:SecondLaw3}).
Indeed, the inequality~(\ref{eq:SecondLaw1}) is bounded by
\begin{align}
\beta Q&\le-(\Delta S_\mathrm{ms}+\Delta I_\mathrm{fb})
\vphantom{I(\rho_{SM}^\mathrm{(fb)})}
\nonumber\\
&
=\Delta S_M-I(\rho_{SM}^\mathrm{(fb)})
\nonumber\\
&\le\Delta S_M,
\vphantom{I(\rho_{SM}^\mathrm{(fb)})}
\label{eq:SecondLaw13}
\end{align}
where $I(\rho_{SM}^\mathrm{(fb)})$ is the quantum mutual information between $S$ and $M$ in the state $\rho_{SM}^\mathrm{(fb)}$, which is for sure non-negative $I(\rho_{SM}^\mathrm{(fb)})\ge0$.
See the caption of Fig.~\ref{fig:DemonMemory} for the definition of the quantum mutual information $I(\rho_{SM})$.

The inequalities in~(\ref{eq:SecondLaw12}) and~(\ref{eq:SecondLaw13}) are valid for the steady cycle with any time $\tau$ for the thermal contact in Step 4.
The inequality~(\ref{eq:SecondLaw1}) is tighter than the previously known inequalities~(\ref{eq:SecondLaw2}) and~(\ref{eq:SecondLaw3}) even in the standard scenario where the working substance $S$ is completely thermalized with $\tau\to\infty$ in Step 4 and the cycle starts with  the thermal equilibrium state $\rho_\mathrm{th}$.

\subsection{Measurement Backaction and Necessity of Feedback Control for Quantum Maxwell Demon}
\label{sec:Necessity}
The inequalities~(\ref{eq:SecondLaw2}) and~(\ref{eq:SecondLaw3}) show that the acquisition of information by measurement, i.e.~positive $I_\mathrm{QC}>0$ in~(\ref{eq:SecondLaw2}) and positive $\Delta S_M>0$ in~(\ref{eq:SecondLaw3}), would allow us to achieve $Q>0$ and to go beyond the standard Clausius inequality $\beta Q\le0$.
The acquired information can be exploited via feedback control.
In the quantum case, however, the necessity of the feedback control to violate the standard Clausius inequality is not immediately obvious.
The backaction of quantum measurement might already lead to the violation of the standard Clausius inequality without feedback control.
The inequality~(\ref{eq:SecondLaw1}) helps us clarify this point: \textit{feedback structure is indeed necessary to violate the standard Clausius inequality}.

\begin{table}[b]
\caption{
Effects of quantum measurements.
Any unital quantum measurement $\{\mathcal{M}_i^\mathrm{U}\}$, therefore any bare quantum measurement $\{\mathcal{M}_i^\mathrm{B}\}$, results in a non-negative $\Delta S_\mathrm{ms}\ge0$~\cite{HolevoBook2019,Sagawa2022}.
Any efficient quantum measurement $\{\mathcal{M}_i^\mathrm{E}\}$, therefore any bare quantum measurement $\{\mathcal{M}_i^\mathrm{B}\}$, yields a non-negative $I_\mathrm{QC}\ge0$~\cite{Ozawa1986}.
In the other cases, these quantities can take both positive and negative values, which is indicated by ``indefinite.''
In particular, $I_\mathrm{QC}$ can be negative by a general quantum measurement~\cite{Funo2013,Naghiloo2018}.
}
\begin{ruledtabular}
\begin{tabular}{ccc}
&
backaction
&
information
\\
&
$\Delta S_\mathrm{ms}$
&
$I_\mathrm{QC}$
\\
\hline
general meas.\ $\mathcal{M}_i$
&
indefinite
&
indefinite
\\
efficient meas.\ $\mathcal{M}_i^\mathrm{E}$
&
indefinite
&
$\geq0$
\\
bare meas.\ $\mathcal{M}_i^\mathrm{B}$
&
$\geq 0$
&
$\geq 0$
\end{tabular}
\end{ruledtabular}
\label{table:backaction}
\end{table}

We first point out that the entropy change by a bare quantum measurement $\{\mathcal{M}_i^\mathrm{B}\}$ of the type (\ref{eq:PureMeas}) is always non-negative 
\begin{equation}
\Delta S_\mathrm{ms}\ge0,
\end{equation}
due to the unitality of the bare quantum measurement $\{\mathcal{M}_i^\mathrm{B}\}$~\cite{HolevoBook2019,Sagawa2022}.
Now, if we do not apply any feedback $U_i=\openone$ after the bare measurement $\mathcal{M}_i^\mathrm{B}$, or if we simply apply some control $U_i=U_0$ irrespective of the outcome $i$ of the bare measurement $\mathcal{M}_i^\mathrm{B}$, we have $\Delta I_\mathrm{fb}=0$, due to the unitary invariance of the von Neumann entropy $S(U_0\rho U_0^\dag)=S(\rho)$.
In this case, the inequality~(\ref{eq:SecondLaw1}) is reduced to $\beta Q\le-\Delta S_\mathrm{ms}\le0$.
This means that the standard Clausius inequality $\beta Q\le0$ cannot be violated solely by the bare quantum measurement without feedback control.
In other words, feedback control $U_i$ depending on the outcome $i$ of the measurement $\mathcal{M}_i^\mathrm{B}$ is \textit{necessary} to violate the  standard Clausius inequality $\beta Q\le0$, by inducing a negative enough $\Delta I_\mathrm{fb}<0$.

A bare quantum measurement $\{\mathcal{M}_i^\mathrm{B}\}$ also yields a non-negative QC-mutual information $I_\mathrm{QC}\ge0$, since it falls into the category of efficient quantum measurement~(\ref{eq:EfficientMeas})~\cite{Ozawa1986}.
See Table~\ref{table:backaction}\@.
Then, the inequality~(\ref{eq:SecondLaw2}) tells us that there is a possibility of violating the standard Clausius inequality $\beta Q\le0$.
This, however, does not imply that feedback control is necessary for the violation of the standard Clausius inequality $\beta Q\le0$.
The inequality~(\ref{eq:SecondLaw2}) is not informative enough to exclude the possibility of violating the standard Clausius inequality $\beta Q\le0$ by the backaction of a bare quantum measurement without feedback control (irrespective of whether the cycle starts with the thermal equilibrium state $\rho_\mathrm{th}$ or with the fixed point $\rho_*$ of the finite-time steady cycle).

If a measurement gives rise to a negative $\Delta S_\mathrm{ms}<0$, the right-hand side of the inequality~(\ref{eq:SecondLaw1}) is positive even without feedback $\Delta I_\mathrm{fb}=0$.
This appears to open the possibility of violating the standard Clausius inequality $\beta Q\le0$ without feedback control.
This, however, should be understood in the following way.
If $\Delta S_\mathrm{ms}<0$, this implies that the measurement is not a bare one.
Then, as argued in Sec.~\ref{sec:Measurement}, the polar decompositions of the measurement operators of the measurement reveal that feedback mechanism is implicitly embedded in the measurement.
In other words, such a measurement process is indistinguishable with a bare measurement followed by feedback control.
Feedback control is hidden there.
The inequality~(\ref{eq:SecondLaw1}) clarifies that such \textit{feedback structure is necessary to violate the standard Clausius inequality $\beta Q\le0$}.
This is the main message of the present paper.

In previous works~\cite{Funo2013,Naghiloo2018}, it is pointed out that a general quantum measurement~(\ref{eq:GeneralMeas}) can yield a negative QC-mutual information $I_\mathrm{QC}<0$, due to its coarse-graining character.
Then, according to the inequality~(\ref{eq:SecondLaw2}), the standard Clausius inequality $\beta Q\le0$ cannot be violated by such a general quantum measurement, even if some additional feedback control is applied after it.
Since the inequality~(\ref{eq:SecondLaw1}) is tighter than the inequality~(\ref{eq:SecondLaw2}) as shown in~(\ref{eq:SecondLaw12}), the right-hand side of the inequality~(\ref{eq:SecondLaw1}) is also negative for a general quantum measurement yielding $I_\mathrm{QC}<0$.
Note that, since
\begin{equation}
I_\mathrm{QC}
=I_\mathrm{ms}-\Delta S_\mathrm{ms}
\ge-\Delta S_\mathrm{ms},
\end{equation}
a general quantum measurement yielding $I_\mathrm{QC}<0$ gives rise to $\Delta S_\mathrm{ms}>0$.
The bound $\beta Q\le-(\Delta S_\mathrm{ms}+\Delta I_\mathrm{fb})\le I_\mathrm{QC}<0$ shown in~(\ref{eq:SecondLaw12}) implies that this positive $\Delta S_\mathrm{ms}>0$ cannot be compensated by any feedback control, and the right-hand side of the inequality~(\ref{eq:SecondLaw1}) is bounded to be negative if $I_\mathrm{QC}<0$.

\section{Example: Two-Level System}
\label{sec:Example}
Let us look at an example.
We consider a two-level system $S$, which has two energy levels $\{\ket{0},\ket{1}\}$ with an energy gap $\varepsilon$.
Its Hamiltonian is given by $H_S=\varepsilon\ket{1}\bra{1}$.

We consider quantum measurements with two outcomes for Step~2 of the protocol.
Since any quantum measurement process is equivalent to a bare measurement $\{\mathcal{M}_i^\mathrm{B}\}$ followed by feedback control $\{\mathcal{U}_i\}$, we restrict the measurement in Step~2 to bare measurements $\{\mathcal{M}_i^\mathrm{B}\}$, without loss of generality.
Feedback control $\{\mathcal{U}_i\}$ is applied after it in Step~3.
We parametrize the measurement operators $\{M_i\}$ of the bare measurement $\{\mathcal{M}_i^\mathrm{B}\}$, which should be positive-semidefinite $M_0,M_1\ge0$ and satisfy the normalization condition $M_0^2+M_1^2=\openone$, as
\begin{gather}
M_0
=
V
\begin{pmatrix}
\sqrt{r}&0
\\
0&\sqrt{s}
\end{pmatrix}
V^\dagger
,
\nonumber
\displaybreak[0]
\\
M_1
=
V
\begin{pmatrix}
\sqrt{1-r}&0
\\
0&\sqrt{1-s}
\end{pmatrix}
V^\dagger
,
\end{gather}
and generate $r$ and $s$ randomly and uniformly over the range $r,s\in[0,1]$, and a unitary $V$ according to the Haar measure.
The unitaries $U$ and $\{U_i\}$ in Steps~1 and~3 are also randomly sampled according to the Haar measures.

The thermalization process $\rme^{\mathcal{L}\tau}$ in Step~4 is modeled by the following generator $\mathcal{L}$ of the GKLS form,
\begin{align}
\mathcal{L}(\rho)
={}&
{-\frac{\rmi}{\hbar}}[H_S,\rho]
\nonumber\\
&
{}-\frac{1}{2}\gamma(1+n_B)
(
L_-^\dag L_-\rho
+
\rho L_-^\dag L_-
-2
L_-\rho L_-^\dag
)
\nonumber\\
&
{}-\frac{1}{2}\gamma n_B
(
L_+^\dag L_+\rho
+
\rho L_+^\dag L_+
-2
L_+\rho L_+^\dag
),
\end{align}
where $L_-=\ket{0}\bra{1}$ and $L_+=\ket{1}\bra{0}$ are jump operators, $\gamma$ is a decay rate, and $n_B=(\rme^{\beta\varepsilon}-1)^{-1}$ is the Bose distribution function.
This thermalizes $S$ as $\rme^{\mathcal{L}\tau}(\rho)\to\rho_\mathrm{th}=\rme^{-\beta H_S}/Z_S$ in the long-time limit $\tau\to\infty$ for any state $\rho$, but we apply this map for a finite time $\tau$.

Once the measurement operators $\{M_i\}$ and the unitaries $U$ and $\{U_i\}$ are sampled, we numerically find the fixed point $\rho_*$ of the thermodynamic cycle $\mathcal{E}=\sum_{i=0,1}\rme^{\mathcal{L}\tau}\circ\mathcal{U}_i\circ\mathcal{M}_i^\mathrm{B}\circ\mathcal{U}$, and evaluate the relevant thermodynamic quantities in the steady cycle.

\begin{figure}[t]
\begin{tabular}{l@{\ }l}
\footnotesize(a)
&
\footnotesize(b)
\\
\includegraphics[height=0.345\linewidth]{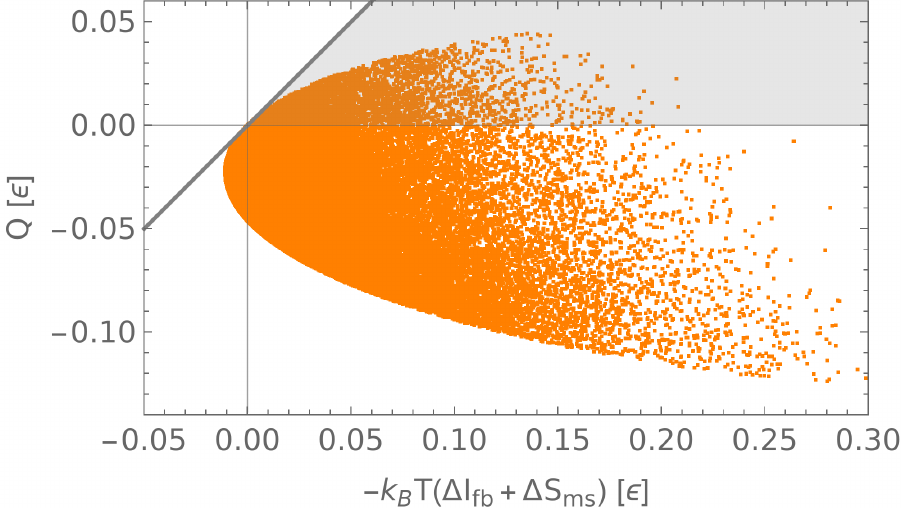}
&
\includegraphics[height=0.345\linewidth]{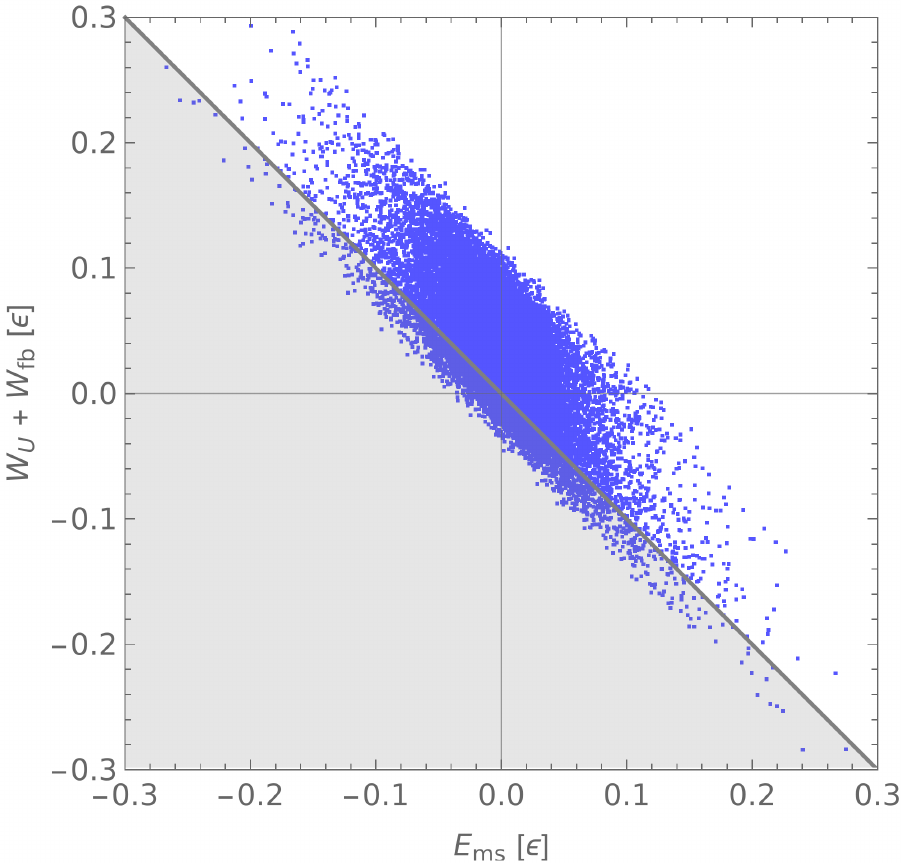}
\end{tabular}
\caption{Thermodynamic quantities in steady cycles for the two-level system $S$. (a) The heat $Q$ absorbed by $S$ from heat bath $B$ and its upper bound $-k_BT(\Delta S_\mathrm{ms}+\Delta I_\mathrm{fb})$ given by the generalized second law~(\ref{eq:SecondLaw1}). The region below the solid line is the region allowed by the generalized second law~(\ref{eq:SecondLaw1}). (b) The work $W=W_U+W_\mathrm{fb}$ done on $S$ by the unitary operation in Step~1 and the feedback control in Step~3 vs the energy gain $E_\mathrm{ms}$ by the backaction of the measurement. 
The measurement operators $\{M_0,M_1\}$ and the unitaries $U$ and $\{U_0,U_1\}$ are randomly sampled, as explained in the text.
The parameters are chosen to be $k_BT=\beta^{-1}=\varepsilon$, $\gamma=0.1\varepsilon/\hbar$, and $\tau=2\hbar/\varepsilon$.
In both panels (a) and (b), the dots in the shaded regions are the samples with $Q=-(W_U+E_\mathrm{ms}+W_\mathrm{fb})>0$, violating the standard Clausius inequality $\beta Q\le0$.}
\label{fig:feedback}
\end{figure}

The heat $Q$ absorbed by the two-level system $S$ from the heat bath $B$ and its upper bound $-k_BT(\Delta S_\mathrm{ms}+\Delta I_\mathrm{fb})$ given by the generalized second law~(\ref{eq:SecondLaw1}) are evaluated for sampled protocols and shown in Fig.~\ref{fig:feedback}(a).
The dots in the shaded region are the samples with $Q>0$.
There are actually cycles that violate the standard Clausius inequality $\beta Q\le0$.

The energy gain $E_\mathrm{ms}$ by the measurement and the work $W=W_U+W_\mathrm{fb}$ done on $S$ by the unitary operation in Step~1 and the feedback control in Step~3 are shown in Fig.~\ref{fig:feedback}(b).
The dots in the shaded region are the samples with $Q=-(W_U+E_\mathrm{ms}+W_\mathrm{fb})>0$, violating the standard Clausius inequality $\beta Q\le0$.
The measurement backaction can induce both $E_\mathrm{ms}>0$ and $E_\mathrm{ms}<0$.
In particular, there are cycles yielding $E_\mathrm{ms}<0$, $W=W_U+W_\mathrm{fb}>0$, and $Q>0$.
In this case, work is not extracted by the controls ($W=W_U+W_\mathrm{fb}>0$), and the absorbed heat ($Q>0$) completely dissipates by the backaction of the measurement ($E_\mathrm{ms}<0$).
This never happens in classical thermodynamic cycles, and this is a characteristic feature in quantum thermodynamics.

\begin{figure}[t]
\begin{tabular}{l@{\ }l}
\footnotesize(a)
&
\footnotesize(b)
\\
\includegraphics[height=0.345\linewidth]{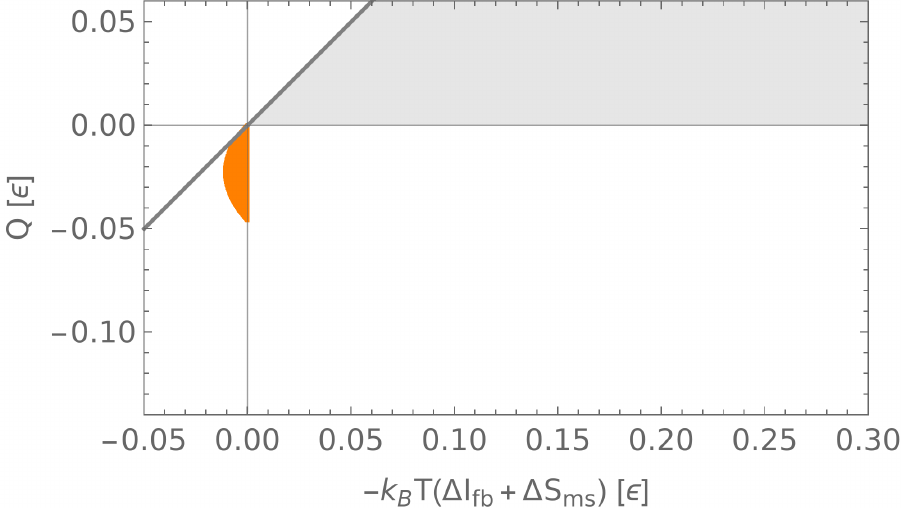}
&
\includegraphics[height=0.345\linewidth]{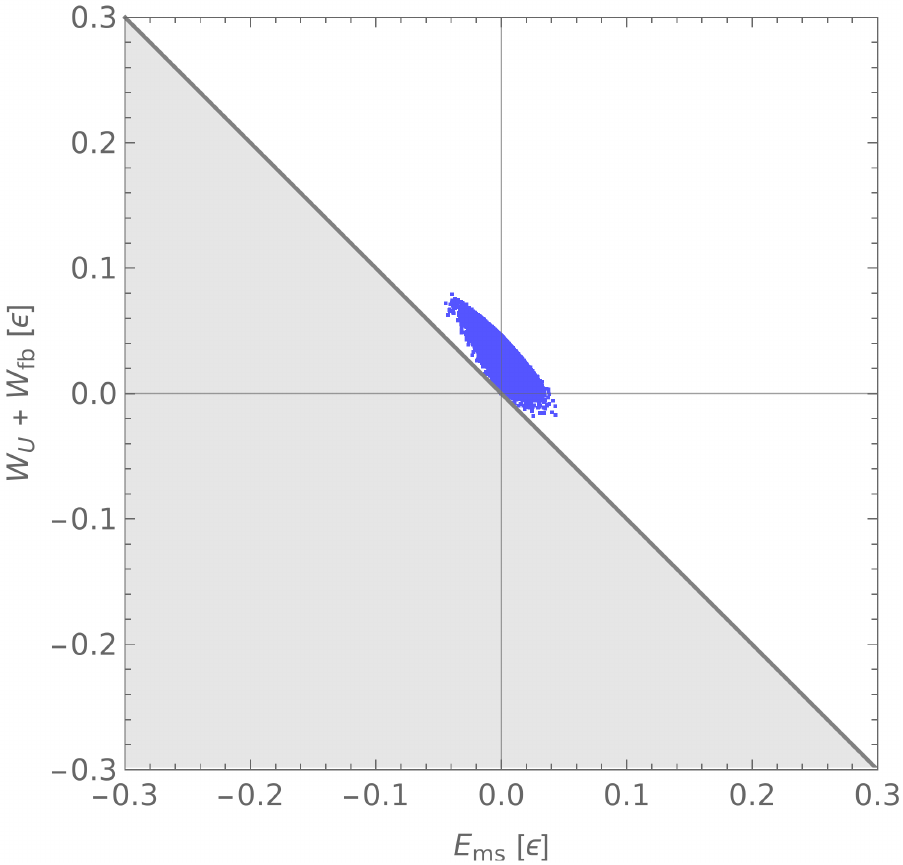}
\end{tabular}
\caption{The same as Fig.~\ref{fig:feedback}, but with the feedback unitary after the measurement is randomly sampled under the constraint $U_0=U_1$ (no feedback).}
\label{fig:nofeedback}
\end{figure}

In Fig.~\ref{fig:nofeedback}, the data obtained by the sampling under the constraint $U_0=U_1$ (with no feedback) are shown.
In this case, there is no cycle that violates the standard Clausius inequality $\beta Q\le0$.
Feedback control is necessary to violate the standard Clausius inequality $\beta Q\le0$.
On the other hand, there are cycles in which work is extracted ($W=W_U+W_\mathrm{fb}<0$) by the controls, even though heat is not supplied ($Q<0$) by heat bath $B$.
This work extracted by the controls, $W=W_U+W_\mathrm{fb}<0$, is supplied by the measurement, $E_\mathrm{ms}>0$.
This is also due to the backaction of the measurement.
This regime ($E_\mathrm{ms}>0$, $Q<0$, and $W<0$ with $U_0=U_1$) is explored in Refs.~\cite{YiTalkerKim2017,Yi2017,YiTalkerKim2018,Anka2021,Lin2021} as a measurement-based heat engine, but with the thermal state $\rho_*=\rho_\mathrm{th}$ in the limit $\tau\to\infty$.

One can also see from Figs.~\ref{fig:feedback}(b) and~\ref{fig:nofeedback}(b) that energy dissipation $E_\mathrm{ms}<0$ by measurement can easily happen (the positivity of $E_\mathrm{ms}>0$ is often argued in the context of measurement-based heat engine~\cite{YiTalkerKim2017,Yi2017,YiTalkerKim2018,Anka2021,Lin2021}).
It remains the case ($E_\mathrm{ms}<0$ can happen) even when each cycle starts with the thermal state $\rho_*=\rho_\mathrm{th}$ in the limit $\tau\to\infty$, because the unitary operation performed in Step~1 breaks the passivity of the thermal state $\rho_\mathrm{th}$.
On the other hand, even in this case, with each cycle starting with the thermal state $\rho_*=\rho_\mathrm{th}$ in the limit $\tau\to\infty$, we are sure about the positivity $W_U+E_\mathrm{ms}\ge0$, since the combined operation $\{\mathcal{M}_i^\mathrm{B}\circ\mathcal{U}\}$ in Steps~1–2 can be regarded as a unital measurement.
See Appendix~\ref{appendix:backaction} for a proof of the passivity of thermal state against unital measurement (see also Ref.~\cite{Anka2021}).
If no feedback is applied in Step~3, namely, if we apply a common unitary in Step~3 irrespective of the outcome $i$ of the measurement, we get $W_U+E_\mathrm{ms}+W_\mathrm{fb}\ge0$, since in this case the combined operation $\{\mathcal{U}_i\circ\mathcal{M}_i^\mathrm{B}\circ\mathcal{U}\}$ in Steps~1–3 as a whole can also be regarded as a unital measurement.
Recalling the first law~(\ref{eq:FirstLaw}), this also proves that the standard Clausius inequality $\beta Q\le0$ holds when the cycle starts and ends with the thermal state $\rho_\mathrm{th}$ with no feedback applied in the cycle.
In the case of the finite-time steady cycle with a finite $\tau$, on the other hand, the proof strategy employed in Appendix~\ref{appendix:backaction} is not useful to get the conclusion that $\beta Q\le0$ holds in the absence of the feedback, due to the presence of the negative last term in~(\ref{eq:EmsDivergence}) of Appendix~\ref{appendix:backaction}\@.

\begin{figure}[tb]
\begin{tabular}{l@{\ }l}
\footnotesize(a) $\tau=0.02\hbar/\varepsilon$
&
\footnotesize(b) $\tau=0.02\hbar/\varepsilon$
\\
\includegraphics[height=0.37\linewidth]{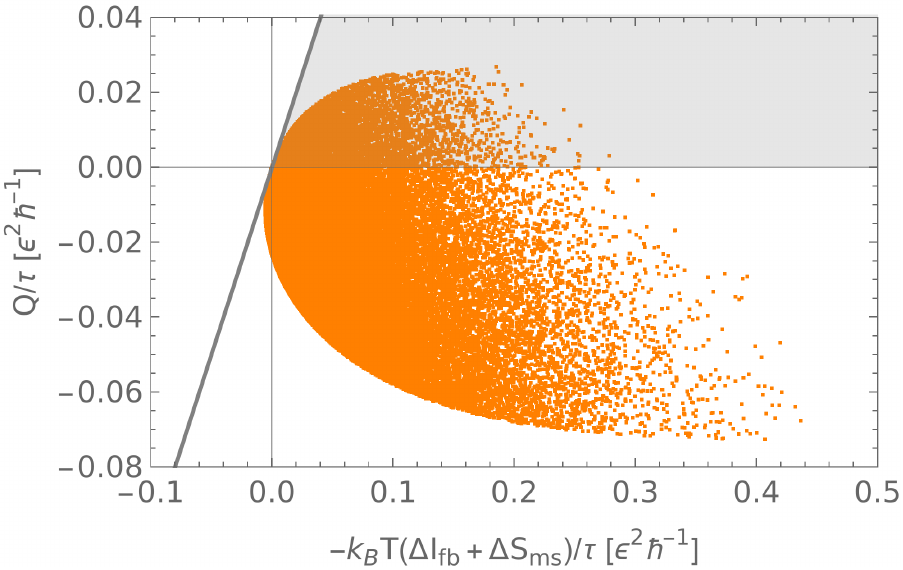}
&
\includegraphics[height=0.37\linewidth]{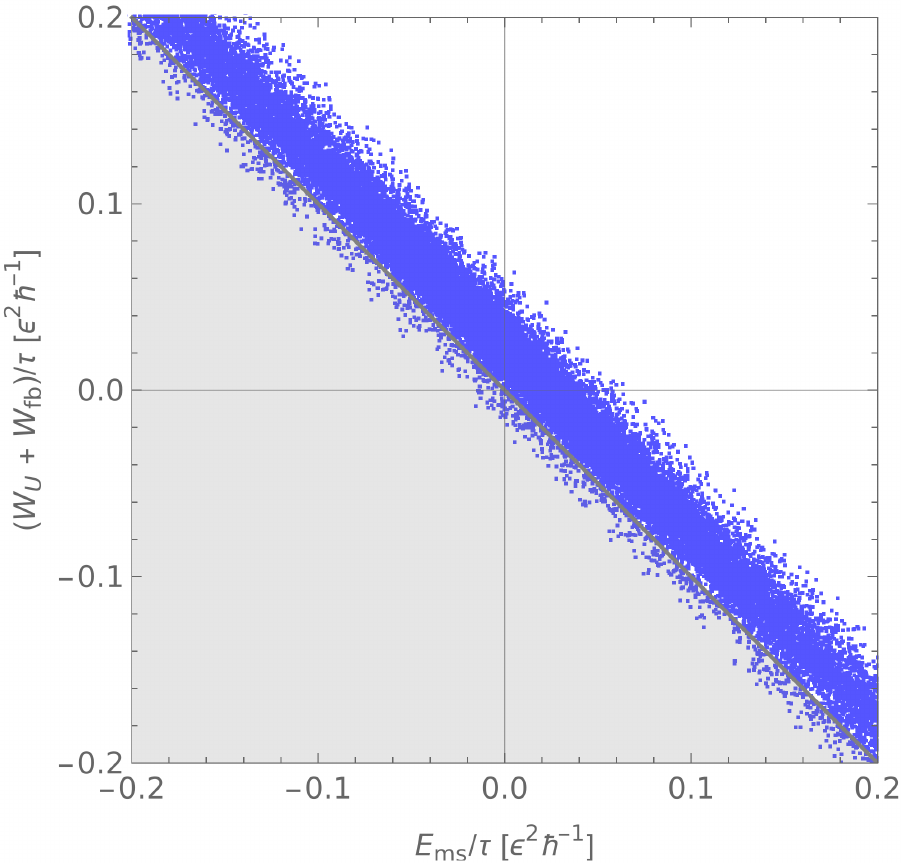}
\\
\footnotesize(c) $\tau=2\hbar/\varepsilon$
&
\footnotesize(d) $\tau=2\hbar/\varepsilon$
\\
\includegraphics[height=0.37\linewidth]{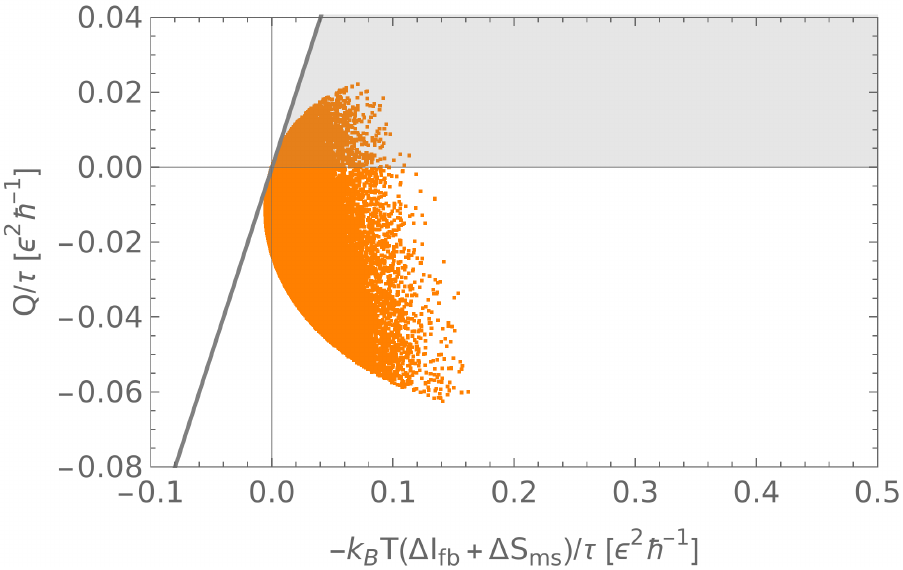}
&
\includegraphics[height=0.37\linewidth]{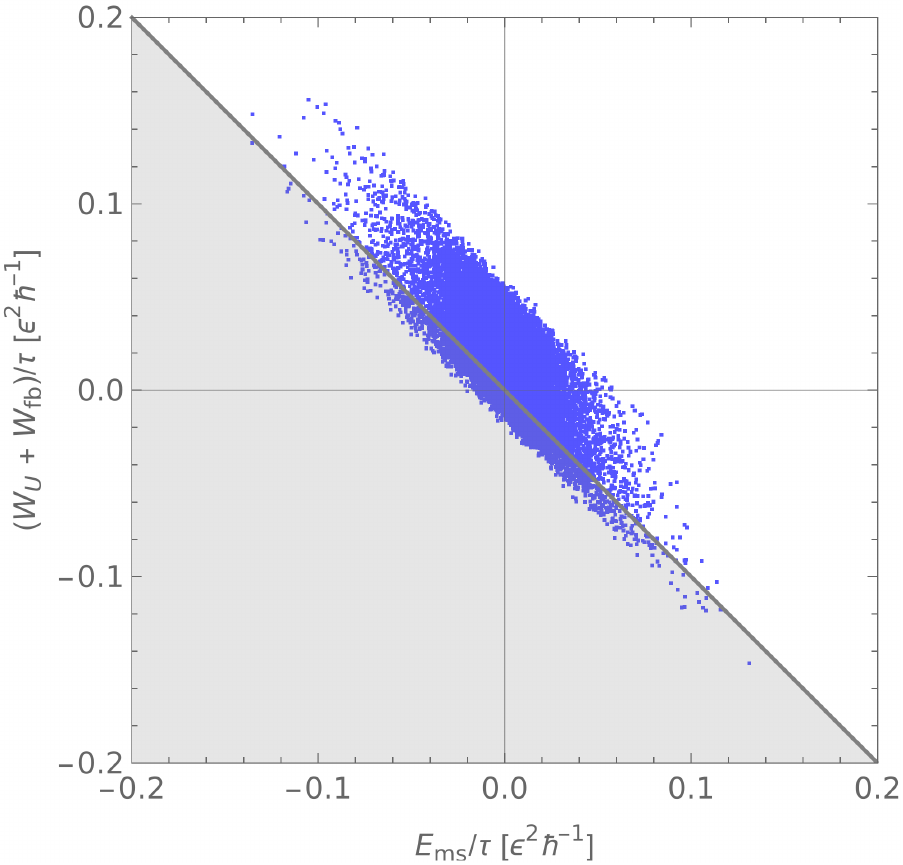}
\\
\footnotesize(e) $\tau=20\hbar/\varepsilon$
&
\footnotesize(f) $\tau=20\hbar/\varepsilon$
\\
\includegraphics[height=0.37\linewidth]{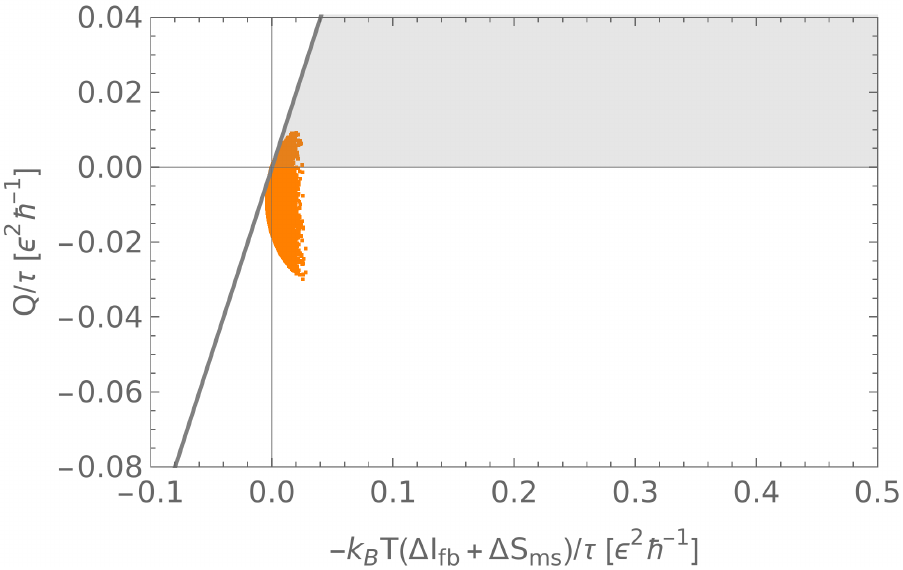}
&
\includegraphics[height=0.37\linewidth]{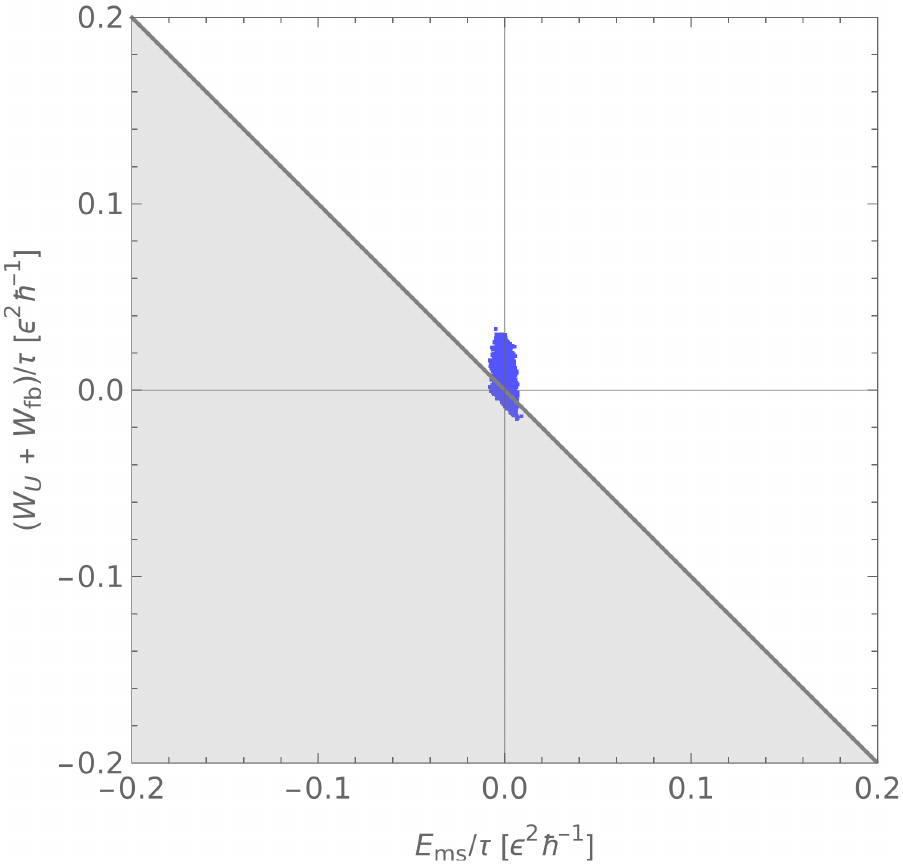}
\\
\end{tabular}
\caption{Thermodynamic quantities per unit time, $Q/\tau_\mathrm{cycle}$, $W/\tau_\mathrm{cycle}=(W_U+W_\mathrm{fb})/\tau_\mathrm{cycle}$, and $E_\mathrm{ms}/\tau_\mathrm{cycle}$, in the steady cycle with period $\tau_\mathrm{cycle}$, for the two-level system $S$. Here, the time spent for $W=W_U+W_\mathrm{fb}$ and $E_\mathrm{ms}$ is assumed to be negligible compared to the time $\tau$ spent for the thermal contact, and hence, $\tau_\mathrm{cycle}\simeq\tau$.  The time $\tau$ for the thermal contact is set at $\tau=0.02\hbar/\varepsilon$ for (a) and (b), $\tau=2\hbar/\varepsilon$ for (c) and (d), and $\tau=20\hbar/\varepsilon$ for (e) and (f). 
The other parameters are $k_BT=\beta^{-1}=\varepsilon$ and $\gamma=0.1\varepsilon/\hbar$.}
\label{fig:rates}
\end{figure}

Let us also look at the powers of the feedback cycle considered here, i.e.~the heat power $Q/\tau_\mathrm{cycle}$ induced by the contact with heat bath $B$ in Step 4, the work power $W/\tau_\mathrm{cycle}=(W_U+W_\mathrm{fb})/\tau_\mathrm{cycle}$ by the unitary controls $\mathcal{U}$ and $\{\mathcal{U}_i\}$ in Steps 1 and 3, and the power $E_\mathrm{ms}/\tau_\mathrm{cycle}$ by the measurement $\{\mathcal{M}_i\}$ in Step 2, in the steady cycle with its period denoted by $\tau_\mathrm{cycle}$.
Note that, if $S$ is fully thermalized with $\tau\to\infty$ in Step~4, these powers vanish.
Finite powers are obtained with a finite cycle period $\tau_\mathrm{cycle}$.
We here focus on the regime where the time spent for the controls $\mathcal{U}$, $\{\mathcal{U}_i\}$, and the measurement $\{\mathcal{M}_i\}$ is negligible compared to the time $\tau$ spent for the thermal contact, and hence, $\tau_\mathrm{cycle}\simeq\tau$.
See Fig.~\ref{fig:rates}.
The maximum powers are enhanced as the time $\tau$ for the thermal contact is reduced.
Higher powers are available by faster cycles, within the regime considered here.

\section{Conclusions}
\label{sec:Conclusions}
We discussed the quantum Maxwell demon in a thermodynamic feedback cycle in the steady-state regime.
We derived a generalized version of the Clausius inequality~(\ref{eq:SecondLaw1}) valid for a finite-time steady feedback cycle.
This allowed us to answer the question raised in Sec.~\ref{sec:Introduction}: \textit{Feedback control is necessary to violate the standard Clausius inequality even in the finite-time steady cycle}.
The backaction of a ``pure'' quantum measurement just disturbs the working substance and increases its entropy.
This should be compensated by a feedback control, exploiting the information acquired by the measurement, to violate the standard Clausius inequality.

In this work, we have just considered a simple feedback cycle with a single heat bath.
Generalization to feedback cycles involving multiple heat baths would be worth exploring.
This would allow us to cover more interesting nonequilibrium situations~\cite{Schaller2011,Strasberg2013,Strasberg2014,Esposito2017,Chida2017,Buffoni2019}.

\acknowledgments
KY acknowledges supports by the Top Global University Project from the Ministry of Education, Culture, Sports, Science and Technology (MEXT), Japan, and by the Grants-in-Aid for Scientific Research (C) (No.~18K03470) and for Fostering Joint International Research (B) (No.~18KK0073) both from the Japan Society for the Promotion of Science (JSPS).

\appendix
\section{Proof of the Generalized Second Law~(\ref{eq:SecondLaw1})}
\label{appendix:proof2ndlaw}
The generalized second law~(\ref{eq:SecondLaw1}) is derived solely from the monotonicity of a quantum relative entropy under the thermalization process in Step~4 of the steady cycle.
Recall that the quantum relative entropy $D(\rho\|\sigma)=\Tr[\rho(\log\rho-\log\sigma)]$ monotonically decreases as $D(\mathcal{E}(\rho)\|\mathcal{E}(\sigma))\le D(\rho\|\sigma)$ under the action of any CPTP map $\mathcal{E}$~\cite{Hayashi2015,HolevoBook2019,Landi2021,Sagawa2022}.
Applying this monotonicity to $\rho=\rho_\mathrm{fb}=\sum_ip_i\rho_\mathrm{fb}^{(i)}$, $\sigma=\rho_\mathrm{th}$, and $\mathcal{E}=\rme^{\mathcal{L}\tau}$, we get
\begin{align}
0\leq{}&
D(\rho_\mathrm{fb}\|\rho_\mathrm{th})
-D(\rme^{\mathcal{L}\tau}(\rho_\mathrm{fb})\|\rme^{\mathcal{L}\tau}(\rho_\mathrm{th}))
\nonumber\\
={}&
D(\rho_\mathrm{fb}\|\rho_\mathrm{th})
-D(\rho_\mathrm{fin}\|\rho_\mathrm{th})
\nonumber\\
={}&
{-S(\rho_\mathrm{fb})}
+\beta\Tr(H_S\rho_\mathrm{fb})
+S(\rho_\mathrm{fin})
-\beta\Tr(H_S\rho_\mathrm{fin})
\nonumber
\displaybreak[0]
\\
={}&
{-[S(\rho_\mathrm{fb})-S(\rho_*)]}
-\beta[\Tr(H_S\rho_*)-\Tr(H_S\rho_\mathrm{fb})]
\nonumber
\displaybreak[0]
\\
={}&
{-}(\Delta I_\mathrm{fb}+\Delta S_\mathrm{ms})-\beta Q,
\end{align}
where we have used the stationarity $\mathcal{L}(\rho_\mathrm{th})=0$ of the thermal state $\rho_\mathrm{th}=\rme^{-\beta H_S}/Z_S$, the cyclicity condition $\rho_\mathrm{fin}=\rho_\mathrm{ini}=\rho_*$, and
\begin{align}
S(\rho_\mathrm{fb})-S(\rho_*) 
={}&S\biggl(
\sum_ip_i\rho_\mathrm{fb}^{(i)}
\biggr)
-\sum_ip_iS(\rho_\mathrm{fb}^{(i)})
\nonumber
\displaybreak[0]
\\
&{}
+\sum_ip_iS(\rho_\mathrm{ms}^{(i)})
-S\biggl(
\sum_ip_i\rho_\mathrm{ms}^{(i)}
\biggr)
\nonumber
\displaybreak[0]
\\
&{}
+S\biggl(
\sum_ip_i\rho_\mathrm{ms}^{(i)}
\biggr)
-S(\rho_U)
\nonumber
\displaybreak[0]
\\
={}&\Delta I_\mathrm{fb}
+\Delta S_\mathrm{ms},
\end{align}
noting $S(\rho_\mathrm{fb}^{(i)})=S(U_i\rho_\mathrm{ms}^{(i)}U_i^\dag)=S(\rho_\mathrm{ms}^{(i)})$ and $S(\rho_U)=S(U\rho_*U^\dag)=S(\rho_*)$.
This proves the generalized second law~(\ref{eq:SecondLaw1}).

\section{Energy Gain by Unital Quantum Measurement on the Thermal State}
\label{appendix:backaction}
The energy gain $E_\mathrm{ms}^\mathrm{U}$ by the backaction of a unital quantum measurement $\{\mathcal{M}_i^\mathrm{U}\}$ performed in the thermal state $\rho_\mathrm{th}$ is always non-negative $E_\mathrm{ms}^\mathrm{U}\ge0$.
Let us prove it in this appendix.

If a general quantum measurement $\{\mathcal{M}_i\}$ is performed in a general state $\rho$, not necessarily thermal, we have
\begin{equation}
E_\mathrm{ms}
=
k_BT[
\Delta S_\mathrm{ms}
+D(\rho_\mathrm{ms}\|\rho_\mathrm{th})
-D(\rho\|\rho_\mathrm{th})],
\label{eq:EmsDivergence}
\end{equation}
where $\rho_\mathrm{ms}=\mathcal{M}(\rho)$ is the average state after the measurement, $E_\mathrm{ms}=\Tr(H_S\rho_\mathrm{ms})-\Tr(H_S\rho)$ and $\Delta S_\mathrm{ms}=S(\rho_\mathrm{ms})-S(\rho)$ are the energy gain and the entropy change by the measurement, respectively, and $\rho_\mathrm{th}=\rme^{-\beta H_S}/Z_S$ is a reference thermal state at an inverse temperature $\beta=(k_BT)^{-1}$.
If the measurement is a unital one $\{\mathcal{M}_i^\mathrm{U}\}$ and if it is performed in the thermal state $\rho=\rho_\mathrm{th}$, the energy gain by the unital measurement $\{\mathcal{M}_i^\mathrm{U}\}$ can be bounded by 
\begin{equation}
E_\mathrm{ms}^\mathrm{U}
=
k_BT[
\Delta S_\mathrm{ms}^\mathrm{U}
+D(\rho_\mathrm{ms}^\mathrm{U}\|\rho_\mathrm{th})
]
\ge
k_BT\,\Delta S_\mathrm{ms}^\mathrm{U}
\ge0,
\end{equation}
since $D(\rho_\mathrm{ms}^\mathrm{U}\|\rho_\mathrm{th})\ge0$, and $\Delta S_\mathrm{ms}^\mathrm{U}\ge0$ by any unial measurement~\cite{HolevoBook2019,Sagawa2022} (superscripts ``U'' have been put to the quantities to emphasize that the quantities are specialized to the case with the unital measurement).

This shows that no energy dissipation $E_\mathrm{ms}^\mathrm{U}<0$ can be induced by a unital quantum measurement $\{\mathcal{M}_i^\mathrm{U}\}$ performed in the Gibbs state $\rho_\mathrm{th}$.
This is due to the passivity of the Gibbs state $\rho_\mathrm{th}$: no energy can be extracted by cyclic unitary operation~\cite{Lenard1978,ThirringBook2002}. 
Note that $\mathcal{M}^\mathrm{U}=\sum_i\mathcal{M}_i^\mathrm{U}$ is a unital map, and any unital evolution can be expressed as a mixture of unitary processes with the unitaries depending on the input state~\cite{note:3}. 
The passivity against the backaction by unital quantum measurement is also discussed in Ref.~\cite{Anka2021}.


\begin{thebibliography}{10}
\bibitem{note:2}
Here, it is assumed that the Hamiltonian of $S$ is brought back to $H_S$ after
  each step of the protocol. Even if this condition is relaxed and the
  Hamiltonian is allowed to be different from $H_S$ after each step, depending
  on the measurement outcome $i$, the sum $W_U+E_\mathrm{ms}+W_\mathrm{fb}$ in
  the first law~(\ref{eq:FirstLaw}) is not affected and the generalized second
  law~(\ref{eq:SecondLaw1}) remains valid, as long as the condition that the
  Hamiltonian of $S$ is back to $H_S$ after Step~3 is fulfilled.
\end{thebibliography}

\end{document}